\documentclass[twocolumn]{aastex63}

\newcommand{\htp}{H$_3^+$}
\newcommand{\htt}{H$_2$}
\newcommand{\water}{H$_2$O}
\newcommand{\angstrom}{\mbox{\normalfont\AA}}

\usepackage{array}
\usepackage{hyperref}
\usepackage{longtable}
\usepackage{rotating, graphicx}
\newcolumntype{P}[1]{>{\centering\arraybackslash}p{#1}}
\newcolumntype{R}[1]{>{\raggedleft\arraybackslash}p{#1}}
\newcolumntype{L}[1]{>{\raggedright\arraybackslash}p{#1}}

\received{May 18, 2022}
\revised{June 06, 2022}
\accepted{June 07, 2022}
\submitjournal{AJ}

\shorttitle{}
\shortauthors{Gibbs et al.}

\begin{document}

\title{Limits on the Auroral Generation of \htp\ in Brown Dwarf and Extrasolar Giant Planet Atmospheres with Keck/NIRSPEC}



\author[0000-0002-9027-4456]{Aidan Gibbs}
\affiliation{Department of Physics \& Astronomy, University of California, Los Angeles, CA 90095, USA}

\author[0000-0002-0176-8973]{Michael P. Fitzgerald}
\affiliation{Department of Physics \& Astronomy, University of California, Los Angeles, CA 90095, USA}

\begin{abstract}

The molecular ion \htp\ is a potentially powerful tracer of the ionospheres and thermal structures of Jovian planets, but has never been detected in a planetary mass object outside of the solar system. Models predict that \htp\ emission driven by EUV flux and solar wind on hot Jupiters, or by powerful aurorae on brown dwarfs, will be between $10^2$ and $10^5\times$ more intense than that of Jupiter. If optimal conditions for the production of emission do exist, the emission may be detectable by current ground-based instruments or in the near future. We present the first search for \htp\ line emission in brown dwarfs with Keck/NIRSPEC $L^\prime$ high-resolution spectroscopy. Additionally, we survey stars hosting giant planets at semi-major axes near $0.1-0.2$ au, which models suggest may be the best planetary targets. No candidate \htp\ emission is found. The limits we place on the emission of \htp\ from brown dwarfs indicates that auroral generation of \htp\ in these environments likely does not linearly scale from the processes found on Jupiter, plausibly due to deeper atmospheric penetration by precipitating auroral electrons. Detection of \htp\ emission in brown dwarfs may be possible with the \em James Webb Space Telescope \em (\em JWST\em), or future thirty-meter class telescopes.

\end{abstract}

\keywords{BDs, Aurorae, Exoplanets, Exoplanet Atmospheres}


\section{Introduction} \label{sec:intro}

A major goal in exoplanet studies is understanding the diversity and evolution of planetary atmospheres, including atmospheric responses to variable stellar environments and loss rates over time. Within this goal, describing the thermodynamic properties and energy budgets that govern atmospheric temperatures, chemistry, and expansion are paramount. As the interface between an atmosphere and its surrounding space, the upper atmospheres of planets, including the thermosphere, ionosphere, and exosphere, are excellent laboratories to probe atmospheric thermodynamics as they are more sensitive to changes in energy flow than lower atmospheric regions. 

In the past decade, planetary upper atmospheres outside our solar system have begun to be probed by spectroscopic observations of hot Jupiters and brown dwarf planetary analogues. Notably, atmospheric expansion and hydrodynamic escape, as well as thermospheric and ionospheric layers, temperatures, and winds, have all been measured or inferred for exoplanets using tracers such as He I and Na I, among others (see \citealt{Vidal-Madjar2011,Spake2018,Seidel2020,Cauley2021,Allart2019}). Recently, observations of ionospheric species have been described as evidence for the possible first detection of an exoplanetary magnetic field \citep{BenJaffel2021}.

A direct and potentially significant tracer of the ionosphere, which has yet to be detected in an extrasolar planetary mass object, is the molecular ion \htp. \htp\ is expected to form readily in the upper atmosphere via the ionization of \htt\ by extreme-ultraviolet radiation (EUV, $10-100$ nm), or by energetic electrons precipitating along field lines during aurorae: 

$$\text{H}_2 + h\nu \rightarrow \text{H}_2^+ + \text{e}^-$$
$$\text{H}_2 + \text{e}^{-*} \rightarrow \text{H}_2^+ + \text{2e}^-$$
$$\text{H}_2 + \text{H}_2^+ \rightarrow \text{H}_3^+ + \text{H}$$

Destruction of \htp\ then occurs by dissociative recombination, or, in the lower thermosphere by reaction with other molecular species, such as hydrocarbons or \water. Thus, \htp\ is expected to only exist in concentration above the homopause where atmospheric mixing is weak and abundances of heavier species are low. In stellar atmospheres, \htp\ is not expected to be abundant except in the coolest, metal poor stellar atmospheres \citep{Tennyson2001}, such that any detection in a stellar environment can be attributed to a planetary ionosphere.

Within the thermospheres and ionospheres of giant planets and brown dwarfs, \htp\ is expected to be a dominant source of radiative cooling that scales with incoming energy flux, sometimes called the ``\htp\ thermostat" \citep{Maillard2011}. Observations of Jupiter's auroral regions have shown that \htp\ rotational-vibrational emission in the near- and mid-infrared effectively balances variable energy input from EUV flux, the solar wind, and particle precipitation. Its role as a coolant is expected to continue for extrasolar giant planets, at least up to high thermospheric temperatures less than $\sim$10,000 K, above which the formation will be suppressed by the dissociation of \htt\ \citep{Koskinen2007}. In our own solar system, observations of \htp\ in Jupiter, Saturn, and Uranus have been a useful diagnostic of upper atmospheric temperatures, densities, wind dynamics, and cooling, as well as magnetosphere-atmosphere interactions (see a review by \citealt{Miller2020}).

Theoretical predictions for \htp\ emission in close-in giant-planet and brown dwarf atmospheres agree that the emission is likely to be orders of magnitude larger than that emitted by Jupiter. This increase can be driven by elevated EUV flux compared to Jupiter in the case of giant exoplanets, or by powerful magnetic fields and associated aurorae for brown dwarfs or exoplanets with strong magnetospheres. While there is consensus that \htp\ emission is an important thermal mechanism, exactly how much stronger emission is to be expected in these environments is uncertain. Early predictions suggested that a planet like $\tau$ Boo b may have an \htp\ emission luminosity upwards of $10^{17}$ W, roughly $10^5 \times$ that of Jupiter \citep{Miller2000}, with most of that energy in concentrated in a few lines. Subsequent models, however, have been conservative, with \citet{Koskinen2007} predicting a maximum total luminosity for any planet of $10^{15}$ W and \citet{Chadney2016} predicting maximum total luminosities of around $10^{16}$ W for a planet like HD 209458 b between 0.1 and 0.2 au from a young, active K or M dwarf. An additional complication is that individual line strengths do not grow linearly with the total intensity as the thermosphere becomes hotter, as more molecular transitions become populated at higher excitation temperatures. Furthermore, none of these models account for auroral contributions to \htp\ generation, which may boost the \htp\ luminosity depending on the auroral energy deposition.

Free-floating brown dwarfs may represent the best environment for successful \htp\ detection as a number of them possess evidence of powerful aurorae ($10^4 \times$ more energetic than Jupiter's) in the form of electron-cyclotron maser (ECM) emission and, in a few cases, synchronous variable Balmer emission \citep{Hallinan2015, Pineda2016}. The auroral current is postulated to be driven by either the break between co-rotating and subrotating magnetic field lines outside the dwarf \citep{Nichols2012}, or by interaction with a brown dwarf satellite \citep{Hallinan2015}. By assuming that the strength of the aurorae corresponds to a linear increase in the number of precipitating electrons, and that the energy of the electrons do not increase significantly, \citealt{Helling2019} have suggested that \htp\ emission will similarly be near or above $10^4 \times$ that of Jupiter ($10^{16}$ - $10^{18}$ W). \citealt{Pineda2017} argues that if brown dwarf aurorae operate analogously to that of Jupiter, then \htp\ $L^\prime$ emission of \htp\ could carry up to $10\times$ more energy than optical Balmer emission that has been associated with aurorae for some nearby brown dwarfs (up to $10^{18}$ W). These works suggest that detection of \htp\ emission from a free-floating brown dwarf with current instruments is plausible.

A number of attempts have previously been made to detect \htp\ emission in hot Jupiters, most of which have focused on observing the fundamental ro-vibrational \htp\ lines in the $L$ band around $4\mu m$ \citep{Brittain2002, Goto2005,Schkolnik2006,Laughlin2008, Lenz2016}. \citet{Schkolnik2006} carried out a survey of six systems hosting extrasolar giant planets (EGPs) with the CSHELL instrument on NASA IRTF and set the lowest detection limit so far of $6.3\times10^{17}$ W for GJ 436 at $1\sigma$ significance. Most recently, \citet{Lenz2016} observed the HD 209458 system with VLT/CRIRES using occultation spectroscopy to attempt to reduce the effect of the stellar background. They achieved a $3\sigma$ detection limit of $5.3\times10^{18}$ W for the Q(3,0) line at 3985.5 nm, but were significantly hampered by poor weather. 

There has been one observational attempt to detect \htp\ emission in brown dwarfs. \citet{Pineda2017} searched for \htp\ lines in $K$ band, medium resolution spectroscopy of brown dwarfs with known ECM radiation using Keck/MOSFIRE. They did not detect any excess infrared emission attributable to \htp, but did not set emission limits for these objects. Within their discussion, they suggest high-resolution $L$-band spectra to confirm their results, as only about $10\%$ of \htp\ emission is expected in $K$, with the remaining energy emitted mostly in $L$ \citep{Neale1996}. Therefore, a high-resolution search for emission lines in the $L$ band to set observational limits is of high interest despite the added challenge of increased thermal background when observing at longer wavelengths.

In this paper, we describe a search for \htp\ emission in 5 brown dwarfs and 5 systems hosting giant planets using Keck/NIRSPEC. We describe our target selection methodology in Section \ref{sec:targets}. Section \ref{sec:obs} outlines our observations of the 10 targets and our data reduction process. In Section \ref{sec:results}, we present results and set upper limits on emission for each of our targets. We interpret our non-detections in the context of atmospheric and magnetospheric models for brown dwarfs and giant planets in Section \ref{sec:diss}, and finally discuss future paths for detection with observatories such as \em JWST\em.

\begin{figure*}[t]
\centering
\includegraphics[width=0.7\textwidth]{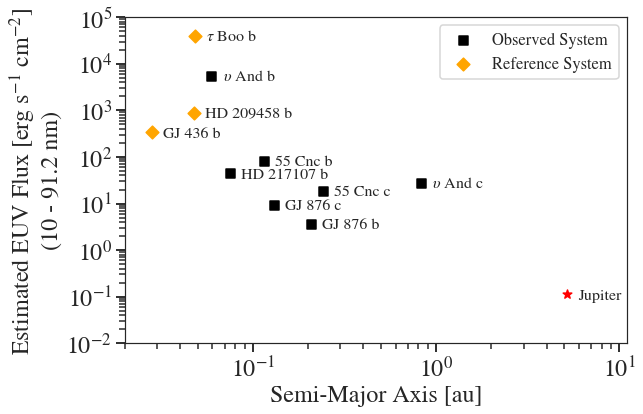}
\caption{\textbf{Estimated EUV flux received by planetary targets.} Giant planets observed in this work are shown by black squares. Orange diamonds mark previously observed giant planets by \citet{Schkolnik2006,Laughlin2008,Lenz2016}, which based on models are likely to have thermospheres too hot to host observable \htp\ emission. EUV data are taken from \citet{Sanz-Forcada2011} and are from observations or coronal modeling. HD 192263 was also observed in this work, but EUV estimates are not available.}
\label{fig:euv_env}
\end{figure*}

\section{Survey Scope and Target Selection} \label{sec:targets}

The primary goal of our study is to attempt to constrain the upper atmospheric and auroral chemistry of brown dwarfs by setting observational emission limits for \htp. As an additional goal, we attempt to detect or set upper limits for \htp\ in giant exoplanets at distances of 0.1 to 0.2 au from solar or later type dwarfs, which models suggest are the most promising planetary targets \citep{Koskinen2007, Chadney2016}. In determining a number of targets to observe, we attempted to balance the integration time per target with sample size such that we would reasonably avoid non-detection due to small number statistics.

To select our brown dwarf targets we relied upon radio observations of ultra-cool dwarfs summarized in \citet{Williams2018}. At present, approximately a dozen brown dwarfs have been observed emitting ECM radiation, which is thought to be produced by auroral processes and is similar to auroral radio emission observed from the poles of Jupiter. We chose the closest brown dwarfs with observed ECM radiation, and which span the spectral range from the end of the main sequence to T dwarfs. \htp\ emission has only been previously modelled in one brown dwarf (2MASSI J1835379+325954) \citep{Helling2019}. Therefore, we decided to select targets with diverse spectral types and therefore atmospheric conditions rather than those that necessarily had the most intense ECM radiation, since the optimal atmospheric conditions for high \htp\ emission in brown dwarfs is still mostly unexplored. We also selected the ultra-cool dwarf binary system 2MASS J07200325-0846499 (Scholz's star) as it is possible the binarity helps stimulate aurorae and \htp\ generation as Io does with Jupiter by producing an interacting plasma \citep{Connerney1993}, although it is not clear if the separation in this system ($a\sim 2.2$\, au, \citealt{Dupuy2019}) is close enough for a similar phenomenon to occur. We do not independently consider the inclination or rotation rates of the dwarfs, although inclinations have been measured for some dwarfs (e.g. \citealt{Vos2017}) and may effect aurorae observability (although this has not been quantified). Our four selected brown dwarf systems can be found in Table \ref{tab:BD}.

The motivation for an additional search for \htp\ in exoplanets is that of three previous observational attempts, two have focused on only one ultra hot Jupiter each ($\tau$ Boo b, \citealt{Laughlin2008}, and HD 209458 b, \citealt{Lenz2016}). While detection of \htp\ in any planet is at the current limits of technical capability, models published after these observations \citep{Chadney2016} suggest that the thermospheres of these specific EGPs are very likely too hot to contain significant \htp\ abundance. Therefore, we decided to include a number of planetary systems that are cooler, yet still have a high ionizing flux from their host star

To select planetary targets, we used the NASA Exoplanet Archive (DOI 10.26133/NEA12). Confirmed planets were sorted based on semi-major axis, bolometric flux received, mass (or $M \sin i$), and distance from Earth. We excluded planets with masses less than that of Saturn, where the thermospheric composition is expected to begin to change from being hydrogen and helium dominated. Planets with semi-major axes less than 0.07 au were excluded based on the probability that their thermospheres are too hot to contain \htp\, through previously described dissociation processes. Otherwise, increased proximity to the star and bolometric flux received were considered to make targets better candidates, as those planets are simply exposed to more ionizing radiation. In reality, it is the EUV, XUV, and X-ray flux received that is most important for \htp\ production, not the bolometric flux, but EUV and XUV are not measured for most exoplanet systems --- due to a lack of observatories in this wavelength region and because of absorption from the interstellar medium --- so the less-informative bolometric flux is used as a proxy. Systems with multiple EGPs were preferred even if some of those giant planets are not necessarily good candidates for detection. Since signal-to-noise of our spectra is of central importance to emission limits, apparent magnitude was used as a primary selection criterion. It is for the same reason that all of our targets are non-transiting systems detected through radial velocity measurements, as known transiting planets farther than 0.1 au from their host stars are still generally much farther from Earth due to the lower probability of having a transiting orbital inclination at increased semi-major axis. The fact that they are non-transiting precludes the possibility of using more sophisticated observational techniques like occultation spectroscopy to search for \htp.

 Our five selected planetary targets can be found in Table \ref{tab:EGP}. A comparison of the estimated EUV flux that some of these planets receive compared to that of Jupiter and a few other exoplanetary systems is shown in Figure \ref{fig:euv_env}. This figure illustrates that we are probing a parameter space between the extremes of thermospheres that are too hot, and too little ionizing EUV radiation to produce observable \htp. EUV flux estimates come from the X-exoplanets database \citep{Sanz-Forcada2011}, which uses coronal modeling to predict EUV if no observational data are available. These values should be viewed as order-of-magnitude. In particular, the EUV flux received by planets around GJ 876 is probably an order-of-magnitude higher as the star is an active M dwarf. 

As a third target class, we chose to observe the white dwarf, brown dwarf (WD-BD) binary GD 1400. This novel target class may combine the benefits of planetary and brown dwarf targets for \htp\ emission. The brown dwarf may have a strong magnetic field and aurorae, but it also receives strong EUV flux from the white dwarf. The white dwarf is dimmer than main-sequence stars in mid-IR, so the system will have a lower stellar contrast. Unfortunately, GD 1400 is very faint despite being the closest WD-BD binary known, and is a thus an extremely challenging target for $L$-band high-resolution spectroscopy.

All together, we selected 10 promising targets for the detection of \htp\ in the atmosphere of a planetary mass object. Highest observational priority was given to the brown dwarf 2MASSI J1835379+325954, as it has strongest evidence for an aurora is the only brown dwarf with a published model for \htp. Otherwise, observing time was based primarily on object observability and scheduling constraints for a given night. Observations of these objects are described in the next section.

\begin{deluxetable*}{R{4.5cm}L{2cm}L{1cm}L{0.5cm}L{0.5cm}L{2cm}L{2cm}}
\tablecaption{Observed Brown Dwarf Sensitivity Limits \label{tab:BD}}
\tablehead{\colhead{Object} & \colhead{Spectral Type}& \colhead{Dist. (pc)} & \colhead{$K$ mag} & \colhead{itime (min)} &
\multicolumn{2}{c}{Emission Limit (W)} \\
\cline{6-7}
\colhead{} & \colhead{} &
\colhead{} & \colhead{} & \colhead{} & \colhead{Q(1,0)} & \colhead{Q(3,0)} }
\startdata
LSPM J0036+1821 (1)  & L3.5 & 8.74 & 11.06 & 68.3 & $6.7\ \times 10^{16}$ & $1.1\ \times 10^{17}$  \\
\hline
SIMP J013656.5+093347.3 (2) & T2 & 6.11 & 12.56 & 53.7 & $2.7\ \times 10^{16}$ & $2.9\ \times 10^{16}$  \\
\hline 
2MASS J07200325-0846499 (3) & M9.5+T5 & 6.02 & 9.47 & 70.0 & $9.3\ \times 10^{16}$ & $5.1\ \times 10^{16}$  \\
\hline
2MASSI J1835379+325954 (4)  & M8.5V & 5.69 & 9.17 & 94.9 & $6.3\ \times 10^{16}$ & $6.7\ \times 10^{16}$ \\
\hline
GD 1400 (5)  & DA4.1 C+L6  & 46.25 & 14.34 & 97.0 & $ 3.5\times 10^{18}$ & $1.6 \times 10^{18}$ \\
\enddata
\tablecomments{Stellar spectral types, distances, and magnitudes come from the SIMBAD astronomical database \citep{Wenger2000}. Radial velocity references: (1) \citet{Blake2010} (2) \citet{Gagne2017} (3) \citet{Burgasser2015} (4) \citet{Deshpande2012} (5) \citet{Napiwotzki2020}. itime refers to the integration time on target not including read times or observing overheads.}
\end{deluxetable*}

\begin{deluxetable*}{R{2.7cm}p{0.1cm}p{0.1cm}p{0.1cm}p{0.1cm}L{2cm}L{2cm}p{1cm}L{1cm}L{1cm}}[t]
\tablecaption{Observed Planetary Target Sensitivity Limits \label{tab:EGP}}

\tablehead{\colhead{Object} & \colhead{Sp. Type} & \colhead{Dist.} & \colhead{$K$ Mag} & \colhead{itime} & \multicolumn{2}{c}{Em. Limit (W)} &  \colhead{Planet} &  \colhead{$M\sin i$} & \colhead{$a$} \\ 
\cline{6-7}
\colhead{} & \colhead{} &
\colhead{(pc)} & \colhead{} & \colhead{(min)} & \colhead{Q(1,0)} & \colhead{Q(3,0)} & \colhead{} & \colhead{(M$_J$)} & \colhead{(au)} } 

\startdata
GJ 876 (1) & M3.5V & 4.68 & 5.01 & 20.5 & $2.2\ \times 10^{17}$  & $6.8 \times 10^{17}$ & c & 0.76 & 0.134 \\
& & & & & $4.7\ \times 10^{17}$ & $5.2\ \times 10^{17}$ & b & 2.39 & 0.214 \\
\hline
HD 192263 (2) & K1/2V & 19.64 & 5.54 & 67.0 & $2.3\ \times 10^{18}$ & $3.2\ \times 10^{18}$ &  b & 0.73 & 0.153 \\
\hline
55 Cancri (3) & K0V & 12.59 & 4.01 & 40.0 & $7.4\ \times 10^{18}$ & $8.4\ \times 10^{18}$ &  b & 0.80 & 0.113 \\
& & & & & $1.6\ \times 10^{18}$ & $1.4\ \times 10^{18}$ & c & 0.16 & 0.237\\
& & & & & & & d & 3.12 & 5.957 \\
\hline
HD 217107 (4) & G8V & 20.07 & 4.54 & 64.2 & $2.9\ \times 10^{18}$ & $8.7\ \times 10^{18}$ & b & 1.39 & 0.075 \\
& & & & & & & c & 4.09 & 5.94 \\
\hline 
$\upsilon$ Andromedae (5) & F9V & 13.49 & 2.86 & 21.2 & $2.9\ \times 10^{18}$ & $5.1\ \times 10^{18}$ &b & 0.69 & 0.059 \\
& &  & & & $2.5\ \times 10^{18}$ & $1.0\ \times 10^{19}$ & c & 13.98 & 0.828 \\
& & & & & & & d & 10.25 & 2.513\\
\hline
\enddata
\tablecomments{Stellar spectral types, distances, and magnitudes come from the SIMBAD astronomical database \citep{Wenger2000}. Stellar radial velocity references: (1) \citet{Fouque2018} (2-5) \citet{Gaia2018}. Planetary ephemeris references: (1) \citet{Millholland2018} (2) \citet{Dragomir2012} (3) \citet{Bourrier2018} (4) \citet{Giovinazzi2020} (5) \citet{Piskorz2017}. itime refers to the integration time on target not including read times or observing overheads.}
\end{deluxetable*}

\section{Observations and Data Reduction} \label{sec:obs}

We observed a total of two and a half nights with Keck/NIRSPEC \citep{McLean1998, Martin2018, Lopez2020} on the 10 targets described in Section \ref{sec:targets}. Individual observations occurred on 2020 August 27th, September 3rd, October 6th, and 2021 March 3rd, UT.

Spectra were obtained using the high-resolution mode with the $KL$ filter. The echelle and cross-disperser angles were set to 62.25$^\circ$ and 33.72$^\circ$ respectively. This setup places six \htp\ lines within order 19 of the echellogram with a wavelength range of approximately 3.94 to 4.02 microns. These are the same lines included in previous searches by \cite{Schkolnik2006} and \citet{Lenz2016}, and which have also been observed prominently in Jupiter (e.g. \citealt{Maillard1990}). Of these six lines, the Q(1,0) and Q(3,0) transitions, at 3953.0 and 3985.5 nm, respectively are expected to be the most intense, with the Q(1,0) being most intense at lower rotational and vibrational temperatures akin to Jupiter ($\sim$500\,K) and Q(3,0) becoming dominant and higher temperatures ($\sim$2000\,K) \citep{Neale1996}. Other transitions in the R and P branch do reach similar intensities as these lines, however, the region around 4 microns is the densest region of prominent lines across low and high excitation temperatures, and other prominent lines could not be observed with a single echelle setting.

Each night, we chose the slit width to be comparable to the average seeing in order to balance target throughput with minimizing sky background. The best seeing we achieved on any night was around 0.4'', so only the 0.432''$\times$12'' (R$\sim25,000$) and wider slits were used. There was no appreciable cloud cover on any night listed above. The precipitable water (PW) ranges for our nights were 1 to 1.2 mm for 27 Aug 2020, 0.6 to 1 mm for 03 Sep 2020, 4 to 6 mm for 06 Oct 2020, and 0.5 to 0.65 mm for 03 Mar 2021.

All observations used a standard ABBA-nodding pattern along the slit that allows removal of thermal background by A-B pair subtraction. At the beginning and end of observing each science target, we observed at least one spectroscopic standard star of A0 or adjacent spectral type, and at similar elevation to the target, for use in correcting telluric lines. Exposure times were chosen for each target to be safely below detector non-linearity at $\sim20,000 $ ADU. Except for the brightest stellar targets, the exposure time is generally limited by the thermal background itself with an exposure of 30 seconds per co-add nearing non-linearity.

Observation parameters for each target are detailed in Tables \ref{tab:BD} and \ref{tab:EGP}. As discussed in Section \ref{sec:targets}, our observation strategy attempted to balance sensitivity with number of targets observed. Thus, no target is observed on sky for longer than $2.5$ hours, corresponding to a maximum integration time of about 90 minutes with readout and nod overheads. Due to target visibility and scheduling constraints, most targets are not observed to this limit.

Initial data reduction steps, including calibration, order rectification, trace fitting, and spectral extraction were carried out using a modified version of the python pipeline described by \citet{Piskorz2016}. Modifications were made to bad pixel correction to prevent clipping with our data. This pipeline uses optimal extraction methods described in \citet{Horne1986}. Stated simply, optimal extraction improves spectral signal-to-noise by weighing pixels based on the wavelength averaged PSF. Reduction steps are carried out individually for each nod pair for bright targets. For dimmer targets, it was sometimes necessary to combine adjacent pairs together, essentially as a co-add, so that the spectral trace could be well-fit. Our faintest target, the WD-BD binary GD 1400, is sufficiently dim that nearly all pairs must be added to confidently fit the trace. 

Raw 1-D science spectra are continuum normalized before being combined into a median spectrum. Continuum normalization is achieved by smoothing the spectrum with a Gaussian filter then constructing a polynomial fit between local maxima, which trace the average continuum level, that is then divided out. This normalization routine changes the systematic structure of the spectra, but not at the scale of the expected \htp\ emission lines. 

Wavelength calibration is performed by a 2nd-order fit of sky emission line positions in the observation data to the same lines in atmospheric transmission models by \citet{Lord1992}. Line centers are determined by a Gaussian fit. We find that even with the spectrograph angles set to the same positions each night, there is still a variation of about 10 angstroms in the wavelength position of a given pixel between nights. The average root-mean-square (RMS) deviation of measured sky line positions compared to our preferred wavelength solutions is 1.1 pixels, with the most accurate wave solution for HD 217107 (RMS$\sim0.37$ pixels) and the least accurate for LSPM J0036+1821 (RMS$\sim1.6$ pixels). For reference, the expected FWHM of any \htp\ line is $\sim4$ pixels.

We use ESO's molecfit tool for telluric correction of the science spectra \citep{Smette2015,Kausch2015}. Molecfit creates a synthetic transmission spectrum using a radiative transfer code and local observatory and meteorological data. Synthetic telluric correction is preferable to using standard stars alone as the synthetic spectrum is theoretically noiseless whereas standard stars necessarily contain photon noise in addition to any systematic errors that can exist in both synthetic and observed spectra. For each target, we generate a synthetic telluric spectrum by fitting a molecfit spectrum to the corresponding standard star. The atmospheric model used is the molecfit default. The exception is that we additionally fit the molecule NO$_2$. The most significant absorption features observed within our wavelength range are caused by NO$_2$. This coincidence is beneficial, as NO$_2$ is expected to have less atmospheric temporal variation compared to other molecules, notably H$_2$O. For planetary targets, we take the additional step of using the best fit telluric model from the standard stars and using it as the start for an additional molecfit refinement using the science spectrum itself. This step is not possible for the brown dwarfs as their intrinsic spectra contain too many absorption lines for molecfit to be able to reliably disentangle the telluric features. An additional benefit of using molecfit that we utilize is a final refinement of the wavelength solution based on the synthetic model. 

As a test for systematic errors that molecfit might introduce to our spectra, we experimented with performing fits using different standard stars for the target GJ 876 (multiple standard stars were observed around this data set). After comparing fits from 3 different standard stars fit separately, there are no systematic differences of large enough magnitude or at a spectral scale to alter the interpretation of the corrected spectrum. Therefore, while some systematic errors are likely introduced by molecfit, we do not believe they could be responsible for an erroneous positive or negative detection.  

For our planetary targets, we expect that the radial velocity does not vary significantly compared to our resolution over our roughly two hour periods of observation. Therefore, we expect the \htp\ line positions to remain effectively stationary so that no spectral shift and add technique is necessary. 

In the next section, we describe our search for \htp\ lines within the data, and our method for setting observational limits for \htp. \newline

\begin{figure*}[t]
\centering
\includegraphics[width=1.0\textwidth]{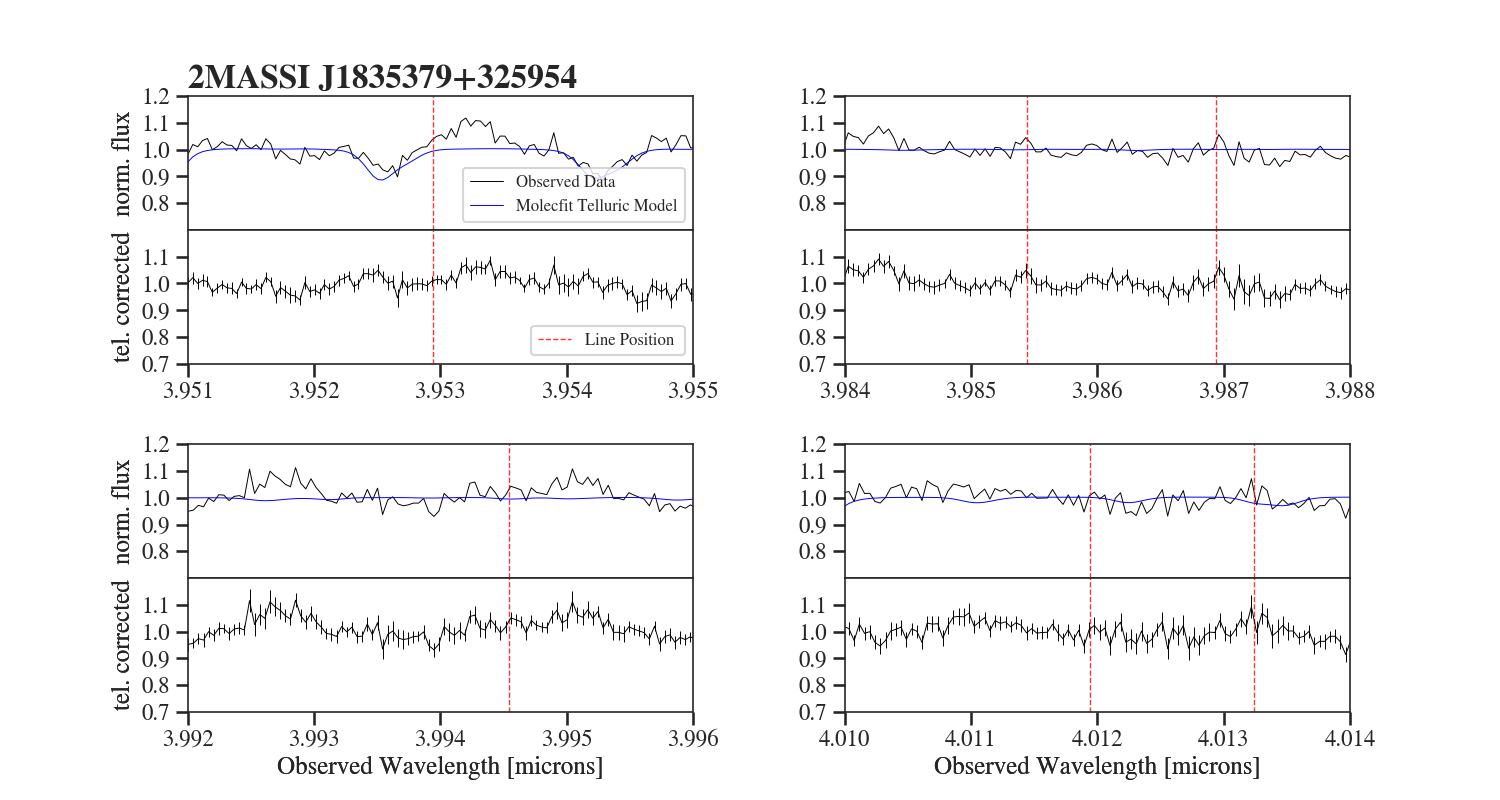}
\caption{\textbf{Spectrum of the free-floating brown dwarf 2MASSI J1835379+325954.} Each panel is zoomed into the expected position of one or more \htp\ emission lines. The top portion of each panel shows the continuum normalized science and telluric model produced by molecfit together, while the bottom portion shows only the science spectrum after the telluric model has been divided out. Expected \htp\ line positions are marked by a red vertical dashed line. The upper limits for \htp\ emission from the Q(1,0) and Q(3,0) lines are $6.3\times 10^{16}$ W and $6.7\times 10^{16}$ W respectively.}
\label{fig:2m1835_spec}
\end{figure*}

\begin{figure*}[t]
\centering
\includegraphics[width=1.0\textwidth]{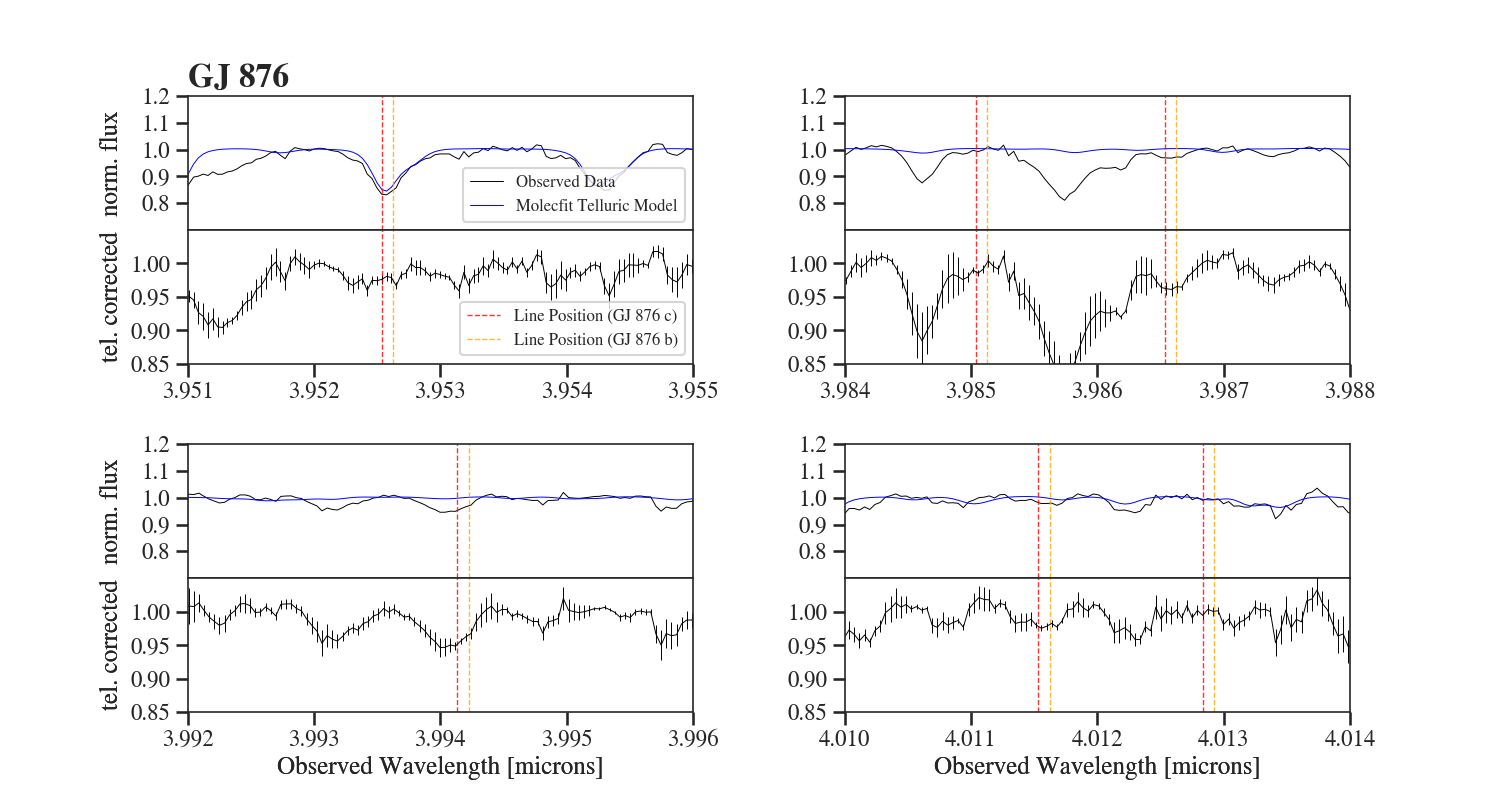}
\caption{\textbf{Spectrum of the M dwarf GJ 876.} This star hosts two giant planets, GJ 876 c, which is closer to the star, and GJ 876 b. Each panel is zoomed into the expected position of one or more \htp\ emission lines. The top portion of each panel shows the continuum normalized science and telluric model produced by molecfit together, while the bottom portion shows only the science spectrum after the telluric model has been divided out. The expected \htp\ line positions are shown separately for each planet. The upper limits for \htp\ emission for GJ 876 c from the Q(1,0) and Q(3,0) lines are $2.2\times 10^{17}$ W and $6.8\times 10^{17}$ W respectively. The corresponding upper limits for GJ 876 b are $4.7\times 10^{17}$ W and $5.2\times 10^{17}$ W.}
\label{fig:gj876_spec}
\end{figure*}

\section{Limits on \htp\ Emission}\label{sec:results}

After data reduction, we search for candidate \htp\ lines within our spectra, and in the absence of any, set limits on the line luminosities that we would have expected to detect given our spectral signal-to-noise.

For all our targets, we elect to take the simple approach of a direct, visual search near the anticipated \htp\ line positions based on the object radial velocities and after correcting for Earth's barycentric velocity. For exoplanetary systems, we also need to account for the radial velocity of each planet, which is dependent on the orbital inclination and phase at observation. Fortunately, most of these exoplanets have relatively well characterized orbits with known inclinations and precise ephemeris data. For example, the uncertainty in the orbital positions of GJ 876 b and c are less than $1\%$. In cases where the inclination of the planet is not known, we simply calculate the range in line positions given a minimum and maximum possible RV at the time of observation. 

Theoretically, we could perform a cross-correlation approach for planetary targets that would be independent of the known planetary RV, either by correlating a \htp\ line model with our combined spectra, or by correlating our spectra at either end of a target's observation sequence to try to catch the slight change in RV of the planet over that time period. We do not believe, however, that either of these options would result in an increased sensitivity. The primary reason is that we only expect a few lines to have intensities that are greater than our spectral noise, and cross-correlation techniques perform best when many lines are available. Also, models of \htp\ line intensities from the ionospheric environment are dependent on non-LTE considerations. If our line models do not account for this correctly it could lead to erroneous results. As for measuring the change in the planetary RV, the maximum change in the RV of any planet over the observation period is on the order of $\sim 1$ km/s, or $\sim0.5 $ \angstrom, which is at the edge of our resolution capability. Finally, by correlating spectra at different points in the observation we necessarily reduce the signal-to-noise of the individual spectra, which makes it less likely that the \htp\ emission will rise above the stellar and thermal background and successfully be correlated.

Spectra for the targets 2MASSI J1835379+325954 and GJ 876 are shown in Figures \ref{fig:2m1835_spec} and \ref{fig:gj876_spec}, with wavelength ranges set to expected line locations. The top panel shows the science and telluric model produced by molecfit before the telluric correction has been applied, and the bottom panel shows the spectrum after the telluric model has been divided out of the science spectrum. Note that while the brown dwarf spectra are not as high signal-to-noise as the stellar spectrum, the observed variability is partly due to closely spaced absorption lines from various species in the brown dwarf atmosphere, and is not purely noise. Spectra for all other targets are provided in appendix figures.

We do not detect any candidate \htp\ emission in any of our brown dwarf or planetary targets. While some targets have a local maxima near one or two line positions, these are all low significance. Specifically, while 2MASSI J1835379+325954 has several low significance maxima at predicted line locations, it does not have any peak at the Q(1,0) line, which is expected to be the second strongest observed line in our wavelength range at a wide range of excitation temperatures.

To calculate emission limits for each target, we follow a similar procedure as in previous searches by \citet{Schkolnik2006} and \citet{Lenz2016}. Our limits are calculated individually for the Q(1,0) and Q(3,0) lines, since these are expected to be the most intense. First, we calculate the standard error of the normalized mean science spectrum in each wavelength bin. Next, if the predicted line location is precisely known (as in the case for brown dwarfs and planets with known inclinations) we calculate both the median of the standard error and the standard deviation of the spectrum within the expected FWHM of an \htp\ line, approximately 4 pixels. To ensure our sensitivity estimate is conservative, we use whichever value is larger as uncertainty in the spectrum at that location. In almost all cases, the standard deviation of the spectrum in the binned wavelength region is higher than our estimate of the standard error from individual exposures within the same region. If the line locations do not have a precise prediction, we calculate the same metrics within the entire region where the line may appear. The theoretical line width of an \htp\ line is calculated from the instrument profile (an output of molecfit) and doppler broadening. Despite the fact that all of our brown dwarf targets are likely rapid rotators (P $<3$ hrs), \htp\ emission from auroral generation is expected to be concentrated towards the poles (as in Jupiter, see latitude profile in \citealt{Drossart2019}) such that at our resolution, the line width is still dominated by the instrument profile. We assume that our synthetic telluric spectrum is noiseless, which is one of the major benefits of using a synthetic model. Next, we estimate the $L^\prime$ magnitude for each target (if no $L^\prime$ photometric measurements are available) using  $K$-$L^\prime$ colors derived from \citep{Bessell1988} for main-sequence stars or by using colors from brown dwarfs with similar spectral types to our targets published in \citet{Golimowski2004}. This estimated $L^\prime$ magnitude, and its associated uncertainty, are then converted to a flux density in the region of our \htp\ lines, and then to a luminosity density given the target parallax. We scale our spectral uncertainty to the luminosity within the line profile and assume that we would follow-up any peak greater than $3\sigma$ significance to get our average line sensitivity for the target.

\begin{figure*}[t]
\centering
\includegraphics[width=1.0\textwidth]{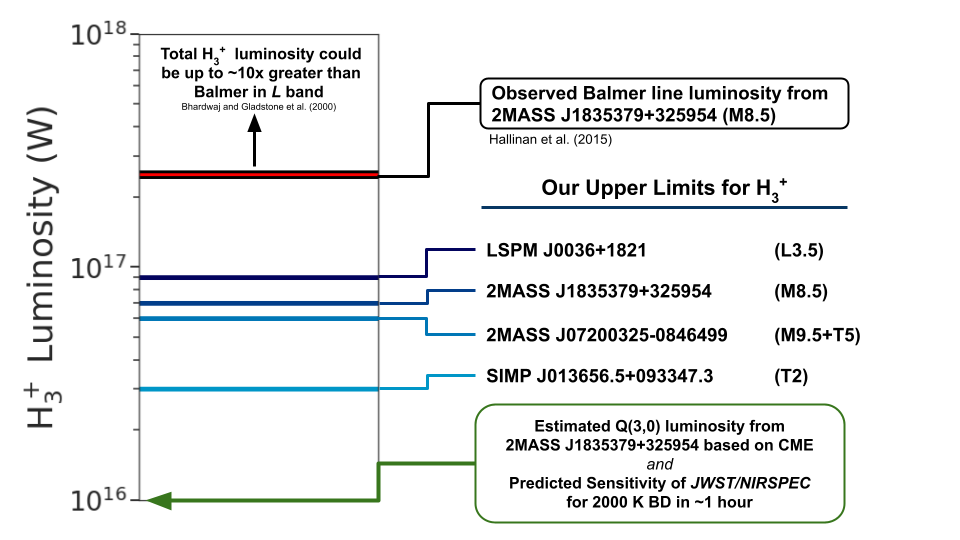}
\caption{\textbf{\htp\ Brown Dwarf Emission Upper Limits.} Blue lines represent sensitivity limits placed in this work for brown dwarfs. The red/black line is the observed luminosity of optical Balmer emission for the brown dwarf 2MASSI J183537+32594 \citep{Hallinan2015} for which total \htp\ emission is thought to have similar or higher luminosity depending on auroral mechanisms \citep{Bhardwaj2000}. The green arrow is our estimated Q(3,0) emission luminosity from 2MASSI J1835379+325954 based on its relative cyclotron maser emission intensity compared to Jupiter and assuming the same auroral electron energy distributions as Jupiter. We estimate that \em JWST/NIRSPEC \em can reach this $10^{16}$ W sensitivity limit to lines around 4 microns for a 2000 K brown dwarf with $\sim 1$ hour of exposure time.}
\label{fig:bd_sensitivity}
\end{figure*}

\begin{figure*}[t]
\centering
\includegraphics[width=1.0\textwidth]{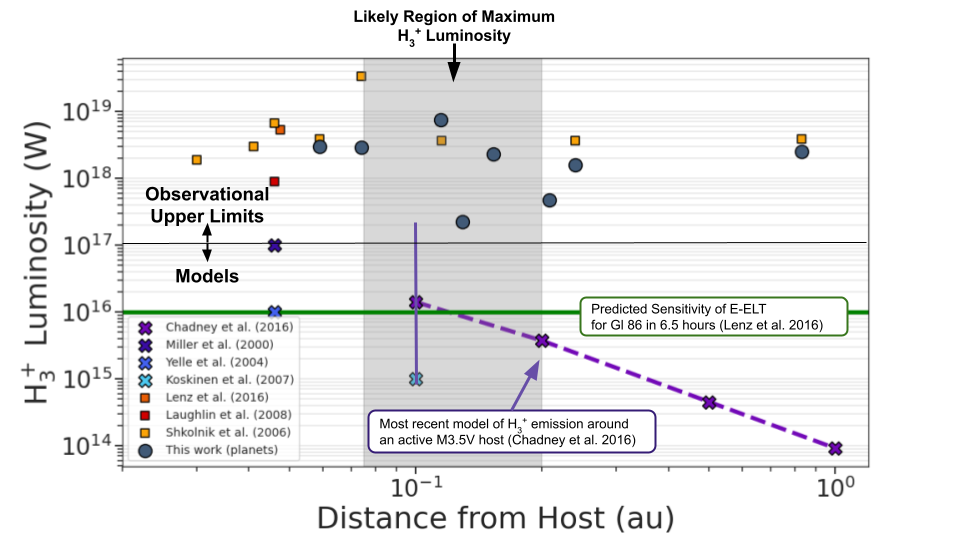}
\caption{\textbf{\htp\ Planetary Emission Sensitivities vs. Theoretical Predictions.} Blue/grey dots represent sensitivity limits placed in this work for extrasolar giant planets. Square points represent sensitivity limits placed by other works. X points represent theoretical predictions for extrasolar giant planets. Note that one observed system can contain multiple giant planets. The green line represents the predicted sensitivity by \citet{Lenz2016} of the E-ELT in 6.5 hours of exposure time using an instrument similar to CRIRES (currently on the VLT) to observe the GJ 86 system. Recent models \citep{Chadney2016} predict that the grey shaded region is the distance from a host star where a planet will reach a maximum \htp\ luminosity for a Sun-like or smaller stellar host.}
\label{fig:sensitivity}
\end{figure*}

Our brown dwarf sensitivities are plotted in Figure \ref{fig:bd_sensitivity}. Planetary sensitivites are shown in Figure \ref{fig:sensitivity} against sensitivities from previous observational attempts and limits from theoretical studies. Brown dwarf sensitivities are roughly an order of magnitude or more stringent than any upper limits previously set for giant exoplanets, which is primarily a result of their low luminosity. We achieve the highest sensitivity for any planetary target so far for GJ 876 c, with a sensitivity of $2.2\ \times10^{17}$ W for the Q(1,0) line. All calculated sensitivities are compiled in Tables \ref{tab:EGP} and \ref{tab:BD}.

\section{Discussion} \label{sec:diss}

In this section, we discuss the physical interpretation of our non-detection for the properties of brown dwarf and giant planet atmospheres and magnetospheres. We will examine brown dwarfs and giant planets separately, beginning with brown dwarfs.

\subsection{Auroral Mechanisms in Brown Dwarf Atmospheres}\label{sec:bd_diss}

Our upper limits for \htp\ emission from brown dwarfs are the lowest limits for any planetary mass objects outside of the solar system by more than an order of magnitude. This reduction in upper limits is possible due to the lower intrinsic luminosity of brown dwarfs compared to bright host stars, and by the relative abundance of nearby brown dwarfs to our own solar system compared to suitable exoplanetary targets.

Importantly, we have corroborated the results of \citet{Pineda2017} that no \htp\ is detected in the atmospheres of free-floating and binary brown dwarfs that are known to possess strong magnetic fields and probable aurorae. Our observations are complementary in that we have observed the wavelength region of the fundamental \htp\ ro-vibrational spectrum at high-resolution in the $L$ band, whereas their study observed the wavelength region of the overtone spectrum at medium-resolution in $K$ band. As \citet{Pineda2017} notes, around 90\% of the \htp\ energy in the Jovian atmosphere is emitted in $L$ band lines while the total $K$ band \htp\ emission in all lines is thought to be at similar luminosities to auroral H$\alpha$ emission. H$\alpha$ emission has been detected and posited to have an auroral origin for 2MASSI J1835379+325954, which was observed in their study and in ours. Figure \ref{fig:bd_sensitivity} shows that the previously observed H$\alpha$ luminosity is well above our upper limits for \htp, however, individual \htp\ line luminosity's in $L$ band may fall below our upper limits by a factor of several depending on excitation temperatures. Nevertheless, the fact that we have not detected \htp\ emission in this target, if the detected H$\alpha$ is indeed auroral, suggests that the auroral processes that generate \htp\ on Jupiter may not be completely analogous to the process that occurs on brown dwarfs. The idea that the physical process is likely to be different is consistent with other observational work by \citet{Saur2018}, who found that the UV spectrum of 2MASSI J1835379+325954 is dissimilar to the UV spectrum from Jupiter's aurora and is at least two orders of magnitude fainter than might be expected from a Jupiter-like auroral mechanism. 

One of the most likely explanations posited by both \citet{Helling2019} and \citet{Pineda2017} is that the electron beams on brown dwarfs have a higher energy distribution than those on Jupiter. Simplistically, one might assume that stronger auorae are due to a greater number of electrons rather than electrons with a different energy distribution. This assumption is justified to an extent because if electron energies are shifted high enough, then the electrons will become relativistic and emit synchrotron radiation rather than ECM, yet strong ECM radiation is still observed from these objects. However, if the electrons do have moderately higher energies (but are not relativistic) they would ionize \htt\ at a greater depth in the brown dwarf atmosphere where \htp\ is quickly destroyed by neutral hydrocarbons or \water\  before it has time to emit. We estimate that our upper limits lack an order of magnitude to sufficiently probe this hypothesis for our targets. If the total \htp\ luminosity of Jupiter \citep{Lam1997} is multiplied by the relative ECM emission of 2MASSI J1835379+325954 compared to Jupiter ($\sim 10^5 \times$), and the rotational and vibrational temperatures of \htp\ are assumed to be 2800\,K, then the Q(3,0) emission luminosity expected from 2MASSI J1835379+325954 is around $10^{16}$ W. The same calculation for our other brown dwarf targets yield similar gaps between our upper limits and the calculated Q(3,0) luminosity, as our targets with lower upper limits all have lower observed ECM radiation intensities. Therefore, an increased sensitivity by a factor of several is required to begin constraining electron distributions to higher energies than that of Jupiter. More than two orders of magnitude decrease in upper limits would be required to fully exclude this hypothesis as individual auroral electrons in a brown dwarf atmosphere could possess $10^2 \times$ more energy than those of Jupiter before becoming relativistic (an increase from $\sim 5$\,keV to $\sim 500$\,keV, \citealt{Gerard2009}). Regardless, if electron energies do increase, \citet{Helling2019} has suggested to instead observe reaction productions of \htp, such as H$_3$O$^+$. At present, there is no calculated H$_3$O$^+$ line list appropriate for an observational attempt. An additional prediction of higher energy electrons is that it will cause more auroral energy to be radiated in the radio, but less in UV and infrared, which could be consistent with surveys of brown dwarfs with radio emission \citep{Gustin2013}. 

Another likely explanation, which is at least of some effect, is that the detection of \htp\ is hampered by the higher thermospheric temperatures of brown dwarfs. Individual line strengths and total luminosity do not scale linearly with temperature, as higher temperatures allow more excitation modes. The thermospheres of our observed brown dwarfs are certainly hotter than that of Jupiter, at least up to the effective temperature of $\sim$2800 K in the case of 2MASSI J1835379+325954 \citep{Berdyugina2017}. This increase in temperature means that although the targeted Q(1,0) and Q(3,0) lines will still be among the most intense, it is no longer valid to assume that those lines will carry most of the \htp\ emission luminosity. Instead it will be carried by many lower intensity lines due to the increase in available excited states. Other explanations are that non-LTE effects are reducing emission intensities or simply that detection requires observing variable aurorae when they are at maximum intensity. Non-LTE effects are certainly important to consider when modelling \htp, however, most models of \htp\ in giant planets have accounted for it and have not shown significant drops in predicted \htp\ luminosity because of it \citep{Koskinen2007, Chadney2016}. While the fact that we have observed multiple targets increases the chance of observing an aurora near a peak of activity, future searches for \htp\ could attempt to simultaneously measure auroral activity by other observables (e.g. ECM radiation) to place \htp\ limits in the context of the activity at the moment of observation.

In future studies, targeting white dwarf - brown dwarf binaries will be an interesting avenue to pursue in addition to free-floating brown dwarfs. Our search for \htp\ from the ground is primarily limited by Earth's atmosphere and background, and observing dim targets ($L>14$ mag) past 2.5 microns is difficult with high-resolution spectroscopy. Our limits for GD 1400 are closer to those of our planetary targets than of our other observed brown dwarfs. Very few WD-BD binaries are known nearby to Earth (GD 1400 being the closest to our knowledge), and realistically it is not possible to target them efficiently with a high-resolution spectroscopic search. WD-BD binaries should be revisited with future-space based observation or with thirty meter telescopes.

\subsection{Extrasolar Giant Planets}\label{sec:egp_diss}

Our sensitivity limits for extrasolar giant planets between 0.1 to 0.2 au from their host stars largely lands us in the same regime as previous observations by \citet{Schkolnik2006}, \citet{Laughlin2008}, and \citet{Lenz2016}. We have achieved the lowest upper limits so far for any extrasolar giant planets with our target GJ 876 by a factor of a few. The value of our study therefore comes primarily from probing a few new systems farther from their stars than have previously been explored. These planets may have cooler thermospheres that allow \htp\ to form. They are also less likely to be tidally locked and therefore may have stronger magnetic fields that could power observable \htp\ emission (although unlikely to be at kilogauss levels). Our observations principally rule out strong auroral processes on these planets that could drive \htp\ production, or if strong aurorae do exist, they are not driving an \htp\ emission increase that might na\"{i}vely by expected, possibly for the same reasons as have been discussed for brown dwarfs. 

\subsection{Prospects for Future \htp\ Detection in Brown Dwarfs}\label{sec:JWST}

Comparing our upper limits for \htp\ emission in brown dwarfs to those of giant exoplanets shows the challenge in dealing with photon noise from a host star. Because of this challenge, future detection has the highest likelihood in brown dwarfs rather than giant exoplanets. Our observations likely set the order-of-magnitude limit achievable with current ground-based instruments for brown dwarfs, at least within one night of observation. Focusing only on one brown dwarf, it is reasonable that luminosity limits of $10^{16}$ W could be achieved with a single night of observations by Keck/NIRSPEC or a similar instrument. At this limit, it may be possible to place lower limits on the energy distribution of precipitating electrons in the brown dwarf atmosphere.

\citet{Lenz2016} previously investigated upper limits that might be achievable for giant planets with the planned E-ELT. Their predicted upper limit of $10^{16}$ W in 6.5 hours of exposure time on the exoplanet system GJ 86 with an instrument like CRIRES on the VLT is shown as part of Figure \ref{fig:sensitivity}. By simple extrapolation, it seems plausible that the E-ELT could reach limits of around $10^{15}$ W in the same amount of time if observing a brown dwarf rather than GJ 86.

In the more near future, it will be possible to observe brown dwarfs in the near-infrared from space using the recently launched \em JWST\em. The greatly reduced thermal background is a significant benefit for potential searches for \htp\ in $L$ and $M$ bands. The drawback is that \em JWST\em/NIRSPEC is medium resolution (R$\sim$2,700), compared to the high-resolution capability of ground-based instruments. This means more blackbody radiation is collected per spectral bin without more \htp\ emission, except for closely spaced lines that are relatively low intensity.

To estimate the ability of \em JWST\em/NIRSPEC to detect \htp\ emission from brown dwarfs, we reverse our steps from Section \ref{sec:results}, going from an emission luminosity to a required signal-to-noise for detection, then use the online \em JWST \em Exposure Time Calculator (ETC) \citep{Pontoppidan2016} to find the time needed for \em JWST \em to reach it. We do these steps using the 2000 K low-temperature PHOENIX model \citep{Phillips2020} provided as a default option within the ETC and determine the needed exposure for three high intensity lines at different wavelengths representative of the range of the fundamental band of \htp. These lines are R(3,-3) at 3533.6 nm, Q(3,0) at 3985.5 nm, and P(6,6) at 4874.4 nm. We find that to reach a limit of $10^{16}$ W requires a total exposure time of around 1 hour for the R(3,-3) and Q(3,0) lines, however, only $\sim 10$ minutes is required to reach that same limit for the P(6,6) line. All 3 of these wavelength regions can be explored simultaneously with the F290LP filter. Thus, despite lower spectral resolution than ground-based spectrographs, \em JWST \em offers significant opportunity to search for \htp\ emission in brown dwarfs and constrain their auroral processes and subsequent atmospheric effects.

\section{Conclusions}\label{sec:conc}

In this study, we have presented a search for \htp\ emission from brown dwarfs with evidence of aurorae and select systems hosting giant planets at semi-major axes of 0.1 to 0.2 au. The key results are as follows.

1) We do not detect any \htp\ emission, but set the first upper limits for \htp\ emission from free-floating brown dwarfs, which, to our knowledge, have never been searched with high-resolution spectroscopy.

2) Our upper limits for \htp\ emission from brown dwarfs range between $2.7$ and $9.3\times 10^{16}$ W, within the upper range of a plausible detection based on models.

3) Our non-detection in brown dwarfs suggests that the aurora-like processes occurring in these objects are probably not analagous to those of Jupiter, namely that electron beams are not able to generate a \htp\ density proportional to the energy of these phenomena. An order-of-magnitude increase in sensitivity is needed to probe this hypothesis further.

4) We set the lowest upper limit for \htp\ emission in a giant exoplanet yet, with a limit of $2.2\times 10^{17}$ W. Despite these limits, we are not able to place additional constraints on giant exoplanet atmospheres or magnetospheres.

5) We suggest that brown dwarfs are the best targets for future \htp\ detection attempts, and show that \em JWST \em will be able to reach emission limits around an order-of-magnitude deeper than current ground-based instruments with equal exposure time.

\acknowledgments

The authors thank Quinn Konopacky for reviewing and providing helpful feedback on the draft of this manuscript.

The data presented herein were obtained at the W. M. Keck Observatory, which is operated as a scientific partnership among the California Institute of Technology, the University of California and the National Aeronautics and Space Administration. The Observatory was made possible by the generous financial support of the W. M. Keck Foundation.

The authors wish to recognize and acknowledge the very significant cultural role and reverence that the summit of Maunakea has always had within the indigenous Hawaiian community.  We are most fortunate to have the opportunity to conduct observations from this mountain.


\facilities{Keck:II (NIRSPEC) }


\software{Numpy \citep{vanderWalt2011}, Pandas\citep {Mckinney2010}, Scipy \citep{Virtanen2019}, Astropy \citep{astropy:2018}}



\clearpage
\appendix

\begin{figure*}[h]
\centering
\includegraphics[width=1.0\textwidth]{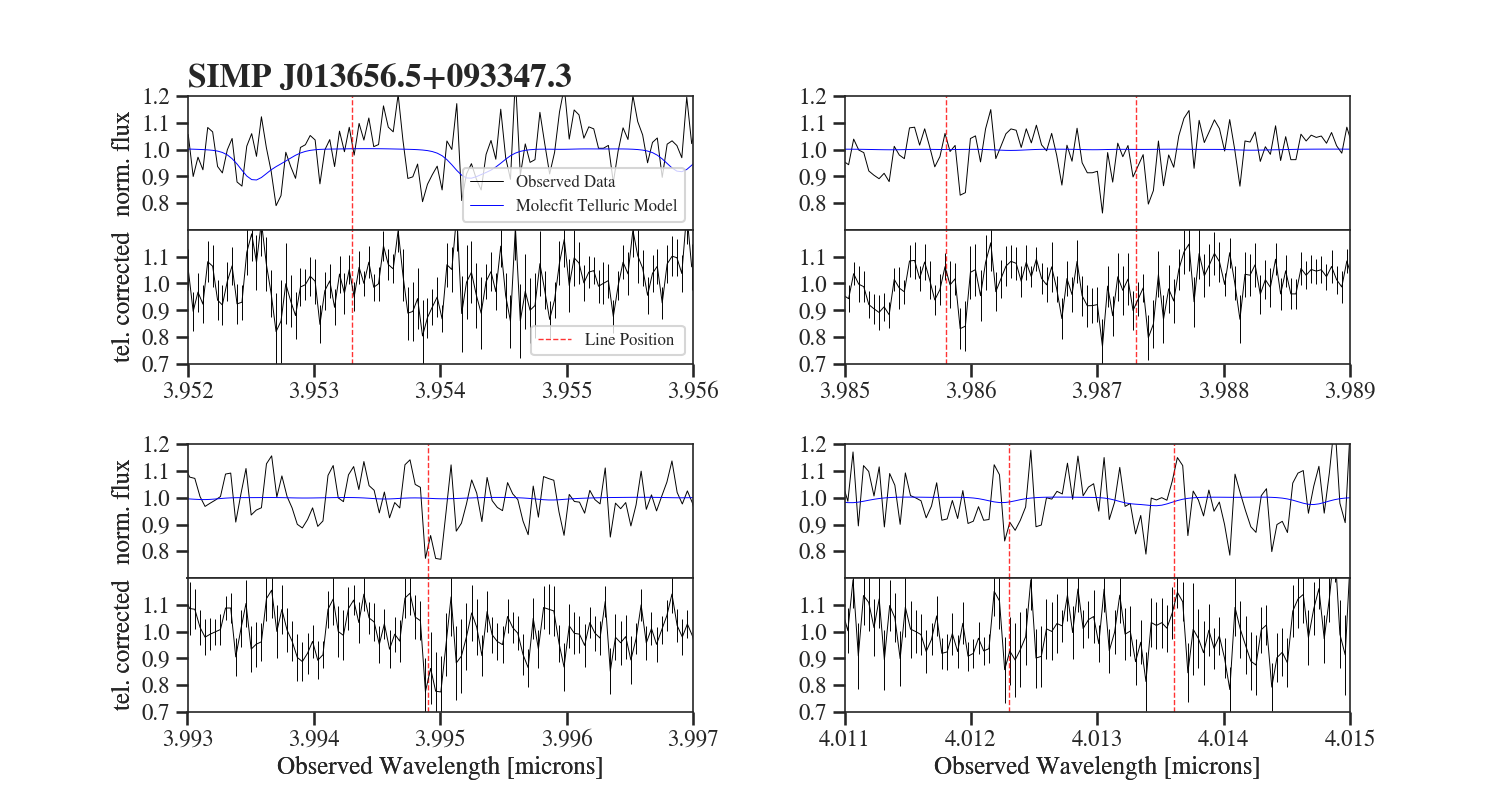}
\caption{\textbf{Spectrum of the free-floating brown dwarf SIMP J013656.5+093347.3.} Each panel is zoomed into the expected position of one or more \htp\ emission lines. The top portion of each panel shows the continuum normalized science and telluric model produced by molecfit together, while the bottom portion shows only the science spectrum after the telluric model has been divided out. Expected \htp\ line positions are marked by a red vertical dashed line. The upper limits for \htp\ emission from the Q(1,0) and Q(3,0) lines are $2.7\times 10^{16}$ W and $2.9\times 10^{16}$ W respectively.}
\label{fig:simp_spec}
\end{figure*}

\begin{figure*}[h]
\centering
\includegraphics[width=1.0\textwidth]{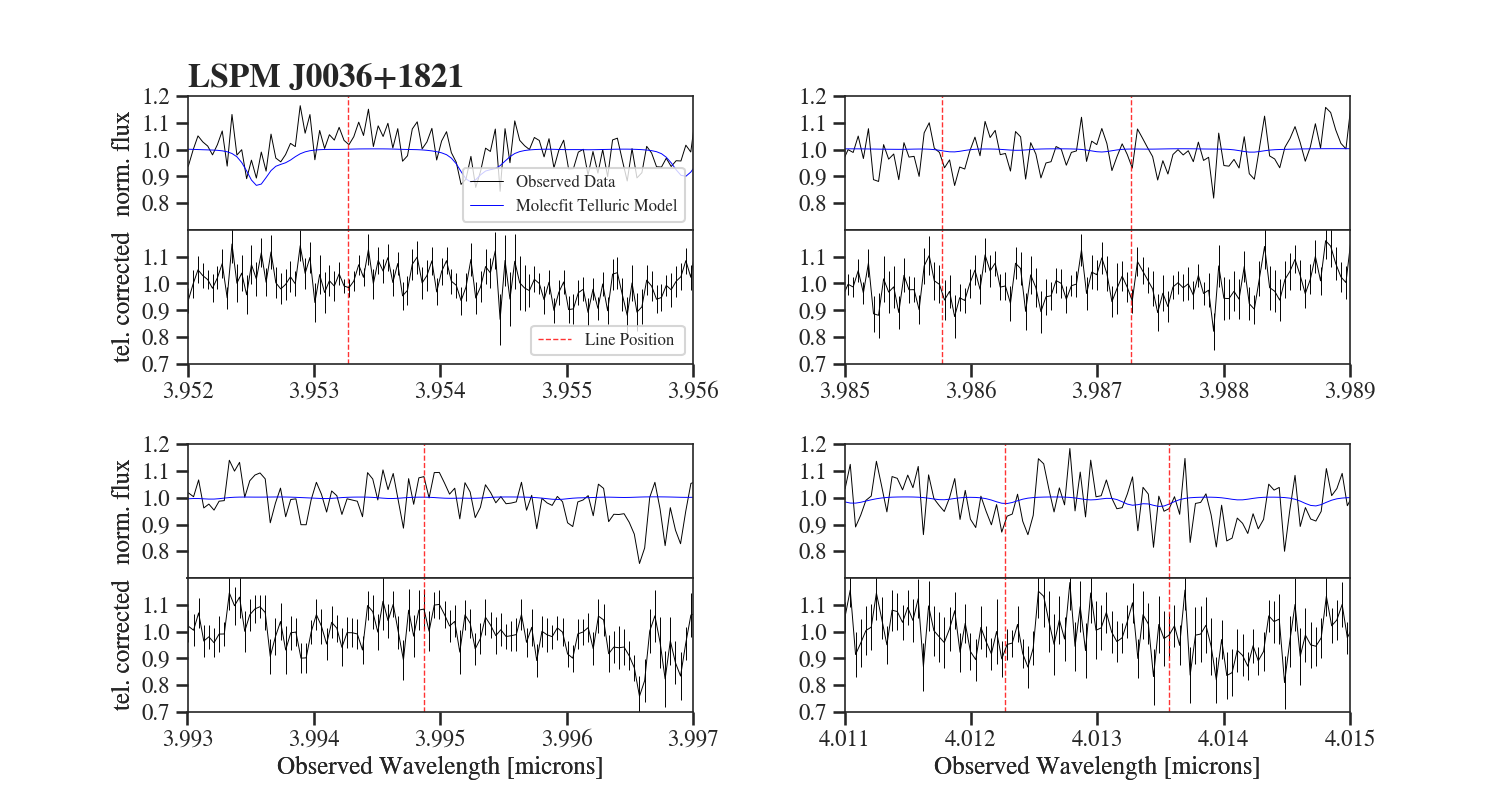}
\caption{\textbf{Spectrum of the free-floating brown dwarf LSPM J0036+1821.} Each panel is zoomed into the expected position of one or more \htp\ emission lines. The top portion of each panel shows the continuum normalized science and telluric model produced by molecfit together, while the bottom portion shows only the science spectrum after the telluric model has been divided out. Expected \htp\ line positions are marked by a red vertical dashed line. The upper limits for \htp\ emission from the Q(1,0) and Q(3,0) lines are $6.7\times 10^{16}$ W and $1.1\times 10^{17}$ W respectively.}
\label{fig:lspm_spec}
\end{figure*}

\begin{figure*}[h]
\centering
\includegraphics[width=1.0\textwidth]{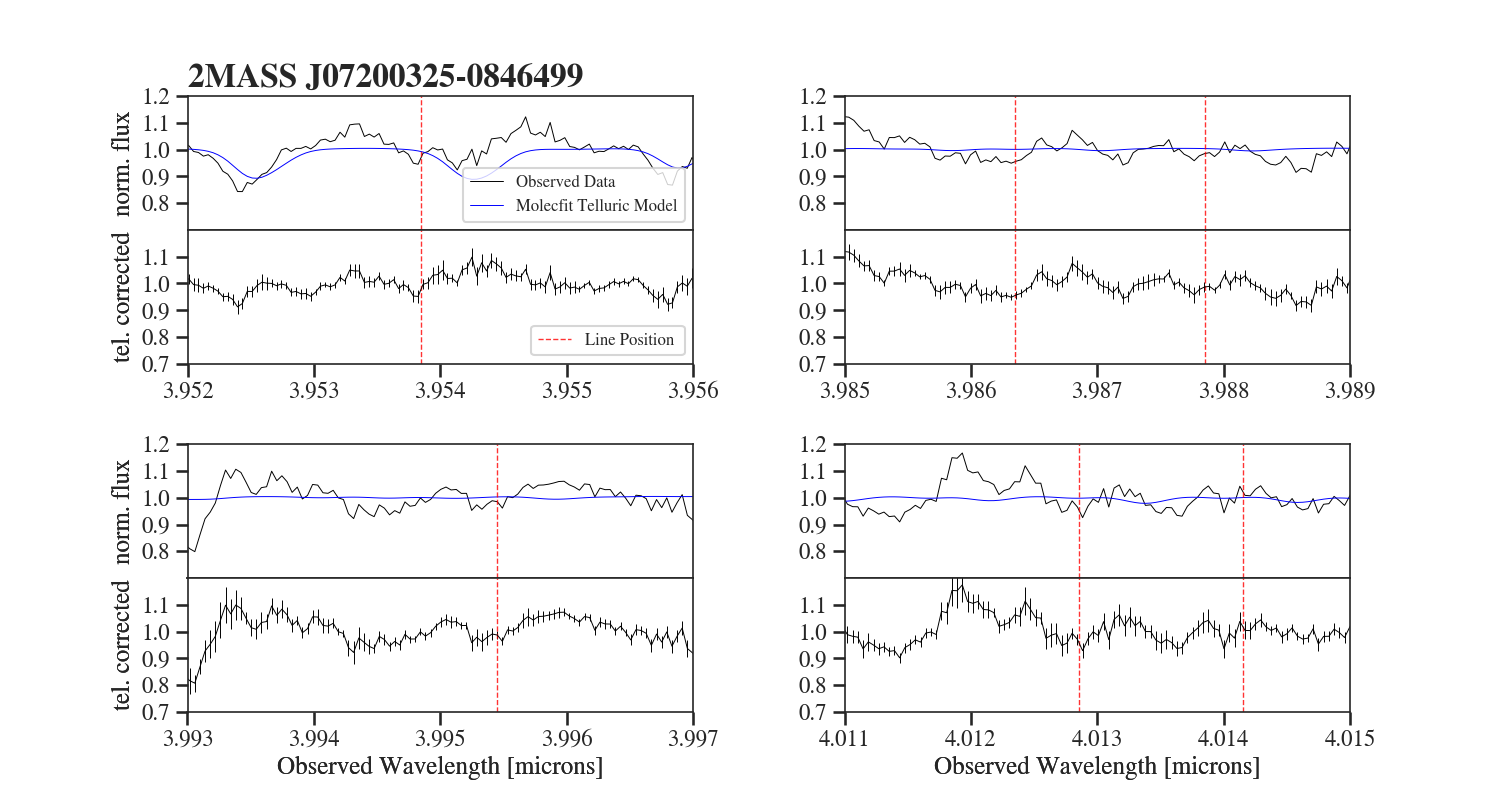}
\caption{\textbf{Spectrum of the binary M dwarf - brown dwarf system 2MASS J07200325-084699.} Each panel is zoomed into the expected position of one or more \htp\ emission lines. The top portion of each panel shows the continuum normalized science and telluric model produced by molecfit together, while the bottom portion shows only the science spectrum after the telluric model has been divided out. Expected \htp\ line positions are marked by a red vertical dashed line. The upper limits for \htp\ emission from the Q(1,0) and Q(3,0) lines are $9.3\times 10^{16}$ W and $5.1\times 10^{16}$ W respectively.}
\label{fig:lspm_spec}
\end{figure*}

\begin{figure*}[t]
\centering
\includegraphics[width=1.0\textwidth]{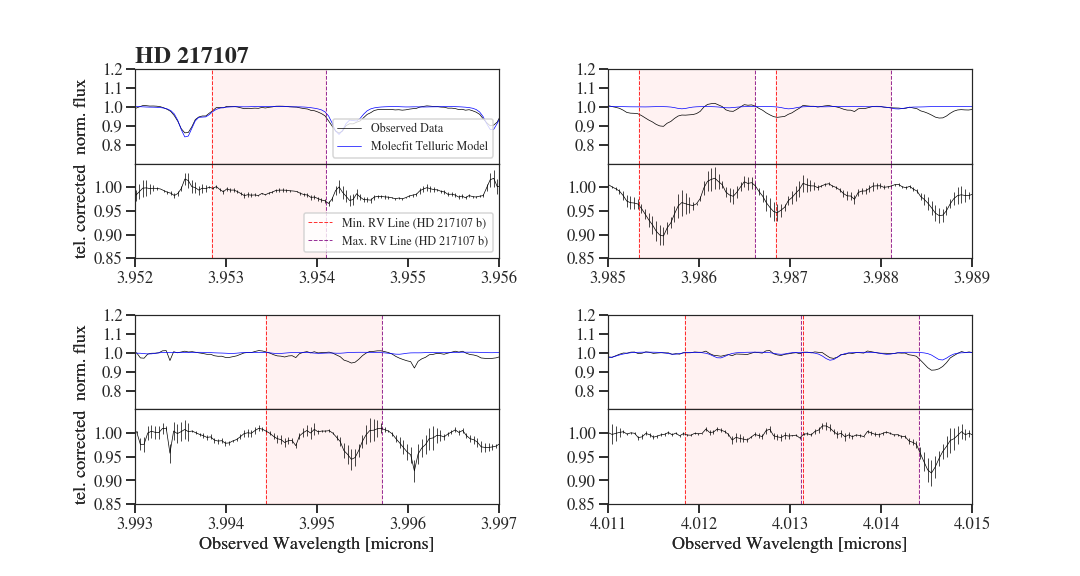}
\caption{\textbf{Spectrum of HD 217107.} This star hosts two giant planets, HD 217107 b and c. The orbit of HD 217107 c is still poorly constrained and is not expected to have observable \htp, so possible line positions are not shown. Each panel is zoomed into the expected position of one or more \htp\ emission lines. The top portion of each panel shows the continuum normalized science and telluric model produced by molecfit together, while the bottom portion shows only the science spectrum after the telluric model has been divided out. The inclination of HD 217107 b is not known, so the line positions show a range from the minimum to maximum possible planetary RV. The upper limits for \htp\ emission for HD 217107 b from the Q(1,0) and Q(3,0) lines are $2.9\times 10^{18}$ W and $8.7\times 10^{18}$ W respectively.}
\label{fig:HD217107_spec}
\end{figure*}

\begin{figure*}[h]
\centering
\includegraphics[width=1.0\textwidth]{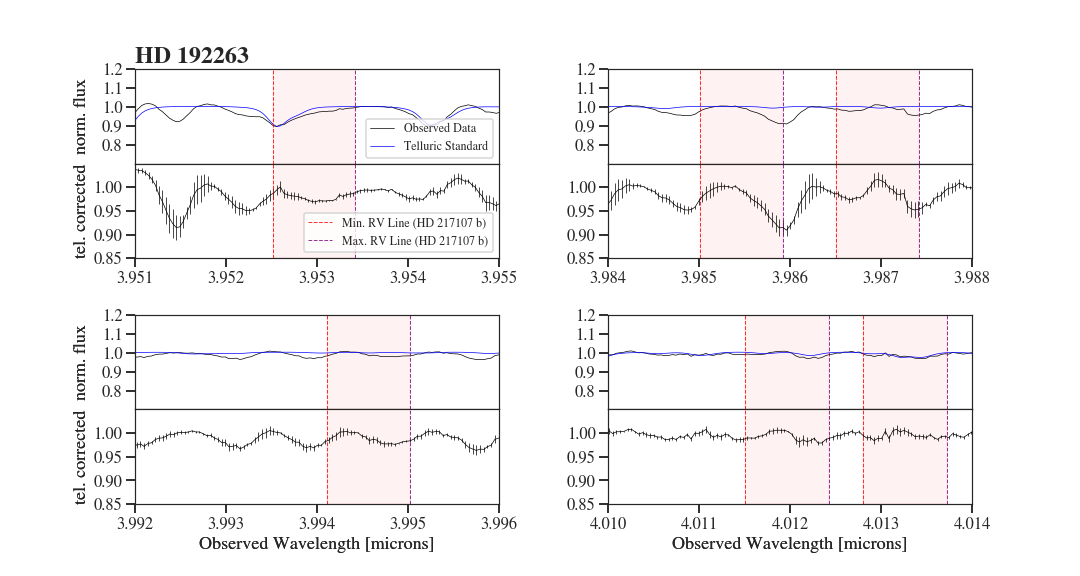}
\caption{\textbf{Spectrum of HD 192263.} This star hosts one giant planets, HD 192263 b. Each panel is zoomed into the expected position of one or more \htp\ emission lines. The top portion of each panel shows the continuum normalized science and telluric model produced by molecfit together, while the bottom portion shows only the science spectrum after the telluric model has been divided out. The inclination of HD 192263 b is not known, so the line positions show a range from the minimum to maximum possible planetary RV. Noise from the telluric spectrum is obviously a limiting factor. In future work, the telluric star will be replaced by a synthetic spectrum. The upper limits for \htp\ emission for HD 192263 from the Q(1,0) and Q(3,0) lines are $2.3\times 10^{18}$ W and $3.2\times 10^{18}$ W respectively.}
\label{fig:HD192263_spec}
\end{figure*}

\begin{figure*}[h]
\centering
\includegraphics[width=1.0\textwidth]{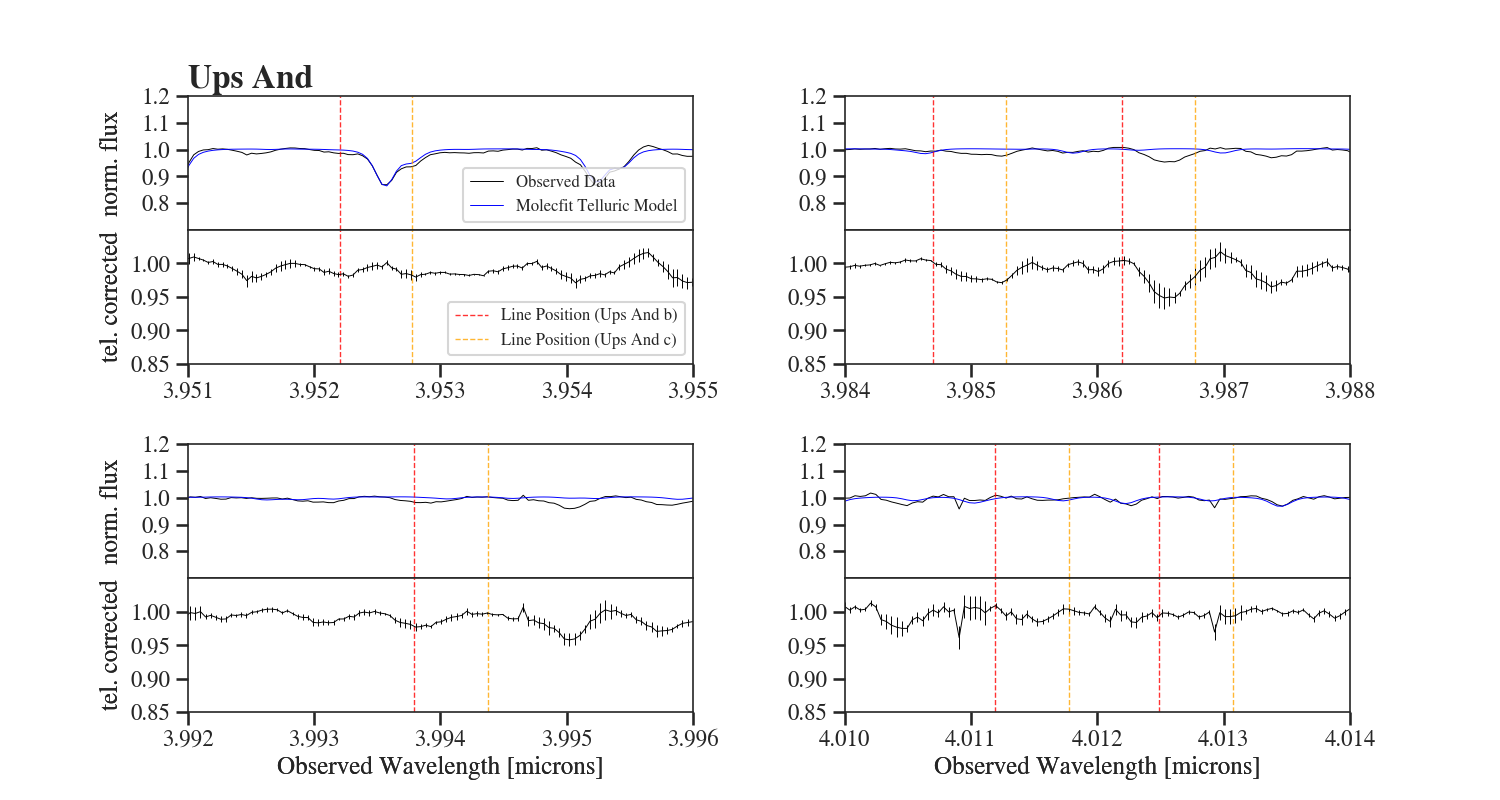}
\caption{\textbf{Spectrum of $\upsilon$ Andromedae.} This star hosts at least three giant planets, $\upsilon$ Andromedae b, c, and d. Each panel is zoomed into the expected position of one or more \htp\ emission lines. The top portion of each panel shows the continuum normalized science and telluric model produced by molecfit together, while the bottom portion shows only the science spectrum after the telluric model has been divided out. The expected \htp\ line positions are shown separately for each planet. $\upsilon$ Andromedae d is not plotted as it is not considered a good candidate for \htp\ detection. The upper limits for \htp\ emission for $\upsilon$ Andromedae b from the Q(1,0) and Q(3,0) lines are $2.9\times 10^{18}$ W and $8.7\times 10^{18}$ W respectively. The corresponding upper limits for $\upsilon$ Andromedae c are $2.5\times 10^{18}$ W and $1.0\times 10^{19}$ W.}
\label{fig:upsand_spec}
\end{figure*}

\begin{figure*}[h]
\centering
\includegraphics[width=1.0\textwidth]{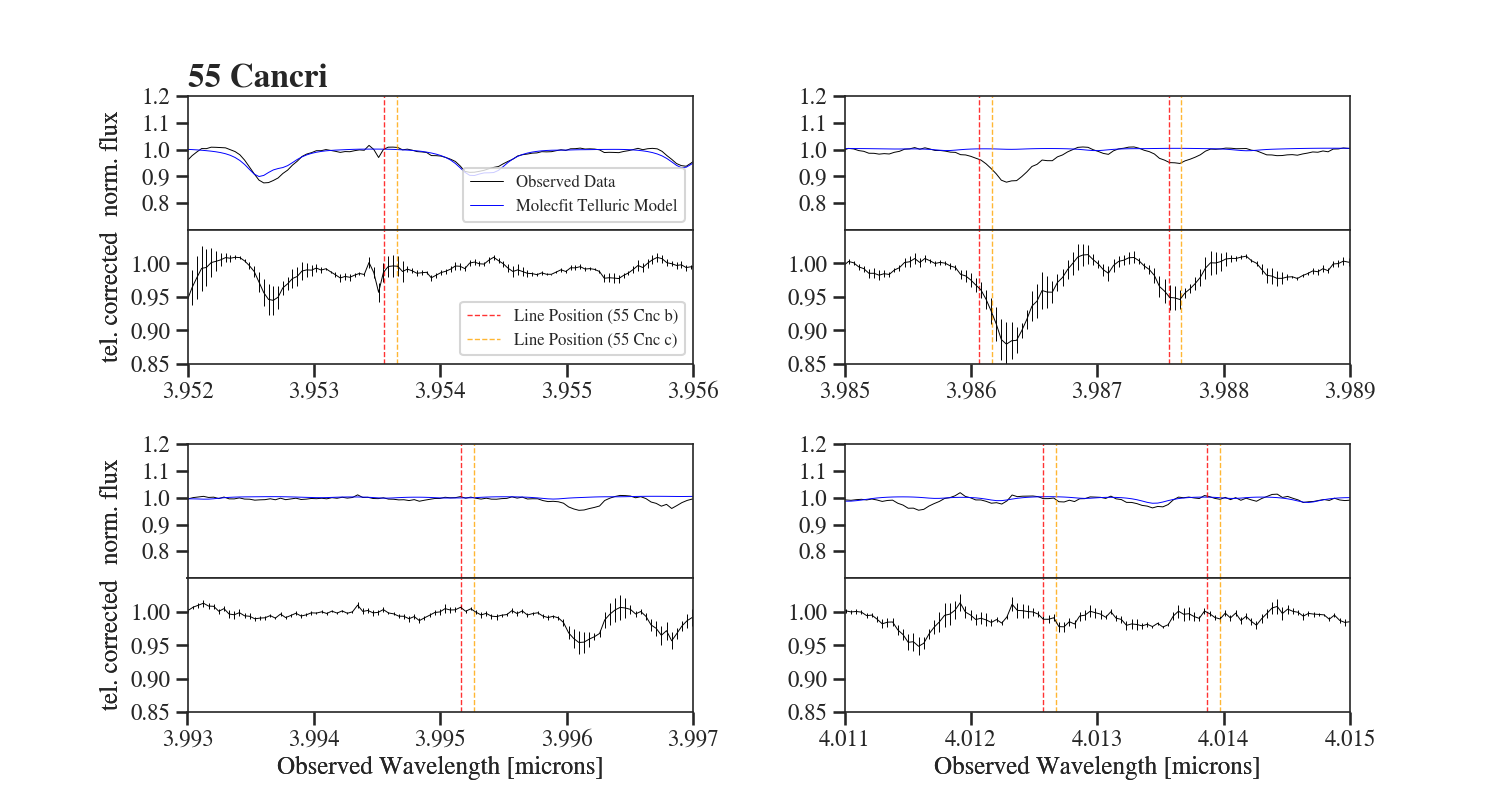}
\caption{\textbf{Spectrum of 55 Cancri.} This star hosts at least four giant planets, 55 Cancri b, c, f, and d. Each panel is zoomed into the expected position of one or more \htp\ emission lines. The top portion of each panel shows the continuum normalized science and telluric model produced by molecfit together, while the bottom portion shows only the science spectrum after the telluric model has been divided out. The expected \htp\ line positions are shown separately for each planet. 55 Cancri d and f are not plotted as they are not considered good candidates for \htp\ detection. The upper limits for \htp\ emission for 55 Cancri b from the Q(1,0) and Q(3,0) lines are $7.4\times 10^{18}$ W and $8.4\times 10^{18}$ W respectively. The corresponding upper limits for 55 Cancri c are $1.6\times 10^{18}$ W and $1.4\times 10^{18}$ W.}
\label{fig:upsand_spec}
\end{figure*}

\clearpage
\bibliography{references}
\bibliographystyle{aasjournal}



\end{document}